\documentstyle[12pt,epsfig]{article}
\newcommand{\bfp}{\mbox{\boldmath $p$}}
\newcommand{\bfP}{\mbox{\boldmath $P$}}
\newcommand{\bfk}{\mbox{\boldmath $k$}}

\def\nostrocostruttino#1\over#2{\mathrel{\mathop{\kern 0pt \rlap
{\hbox{$#1$}}} \hbox{\kern-.135em $#2$}}}
\def\sumint{\nostrocostruttino \sum \over {\displaystyle\int}}
\newcommand{\NP}[1]{{\it Nucl.\ Phys.}\ {\bf #1}}

\newcommand{\PL}[1]{{\it Phys.\ Lett.}\ {\bf #1}}
\newcommand{\PR}[1]{{\it Phys.\ Rev.}\ {\bf #1}}
\newcommand{\PRL}[1]{{\it Phys.\ Rev.\ Lett.}\ {\bf #1}}

\newcommand{\EPJ}[1]{{\it Eur.\ Phys.\ J.}\ {\bf #1}}
\newcommand{\beq}{\begin{equation}}
\newcommand{\eeq}{\end{equation}}
\newcommand{\barr}{\begin{eqnarray}}
\newcommand{\earr}{\end{eqnarray}}
\newcommand{\ba}{\begin{array}}
\newcommand{\ea}{\end{array}}
\newcommand{\pup}{p^\uparrow}
\newcommand{\pdown}{p^\downarrow}
\newcommand{\qup}{q^\uparrow}
\newcommand{\qdown}{q^\downarrow}
\newcommand{\aup}{a^\uparrow}

\newcommand{\bup}{b^\uparrow}
\newcommand{\bdown}{b^\downarrow}
\newcommand{\cp}{c^\uparrow}
\newcommand{\cdown}{c^\downarrow}
\newcommand{\hup}{h^\uparrow}
\newcommand{\hdown}{h^\downarrow}

\newcommand{\la}{\lambda}

\newcommand{\gsim}{\raisebox{-0.07cm}{$\, \stackrel{>}{{\scriptstyle
\sim}}\, $}}

\begin{document}

\begin{flushright}
DFTT 18/2000\\
INFNCA-TH0008 \\
VUTH 00-13 \\
hep-ph/0005081 \\
\end{flushright}

\vskip 0.3cm

\renewcommand{\thefootnote}{\fnsymbol{footnote}}

\begin{center}
{\bf Phenomenology of transverse single spin \\
 asymmetries in inclusive processes\footnote{Talk delivered by
M. Anselmino at the Fifth Workshop on 
Quantum Chromodynamics, January 3-7, 2000, Villefranche, France.} 
\\ }

\vspace{0.4cm}
{\sf M. Anselmino$^1$, M. Boglione$^2$ and F. Murgia$^3$} \\
\vspace*{0.4cm}
{$^1$Dipartimento di Fisica Teorica, Universit\`a di Torino and \\
      INFN, Sezione di Torino, Via P. Giuria 1, 10125 Torino, Italy \\
\vskip 0.5cm
$^2$Dept. of Physics and Astronomy, Vrije Universiteit Amsterdam, \\
        De Boelelaan 1081, 1081 HV Amsterdam, The Netherlands\\
\vskip 0.5cm 
$^3$Dipartimento di Fisica, Universit\`a di Cagliari and \\
INFN, Sezione di Cagliari, CP 170, I-09042 Monserrato (CA), Italy} \\

\vspace*{0.5cm}
\end{center}

\begin{abstract}\noindent
A phenomenological description of single transverse spin asymmetries 
for the inclusive production of hadrons in $p-p$ and $\ell-p$ processes   
is discussed within pQCD and a straightforward generalization of the 
factorization theorem with the inclusion of parton intrinsic transverse 
motion. Fits to existing data, predictions for new processes and 
interpretation of recent results are presented.
\end{abstract}

\vskip 12pt
\noindent
{\bf QCD formalism for {\mbox{\boldmath $A^\uparrow \!\! B \to C X$}} 
at leading twist, {\mbox{\boldmath $\bfk_\perp=0: \> A_N = 0$}}}
\vskip 6pt
It is well known that perturbative QCD and the factorization theorem 
at leading twist \cite{col1,col2} can be used to describe the large $p_T$ 
production of a hadron $C$ resulting from the interaction of two 
polarized hadrons $A$ and $B$:
\barr
\frac{E_C \, d^3\sigma^{A,S_A + B,S_B \to C + X}} {d^{3} \bfp_C} &=&
\sum_{a,b,c,d;\{\lambda\}}
\rho_{\la^{\,}_a, \la^{\prime}_a}^{a/A,S_A} \, f_{a/A}(x_a) \,
\otimes \rho_{\la^{\,}_b, \la^{\prime}_b}^{b/B,S_B} \, f_{b/B}(x_b) 
\label{dsabpol} \\ 
&\otimes& \hat M_{\la^{\,}_c, \la^{\,}_d; \la^{\,}_a, \la^{\,}_b} \,
\hat M^*_{\la^{\prime}_c, \la^{\,}_d; \la^{\prime}_a, \la^{\prime}_b} \,
\otimes D_{\la^{\,}_C,\la^{\,}_C}^{\la^{\,}_c,\la^{\prime}_c}(z) \>, 
\nonumber 
\earr
where $\otimes$ denotes the usual convolutions [see, for example, 
Ref. \cite{noi1} for details]. 
$\rho_{\la^{\,}_a, \la^{\prime}_a}^{a/A,S_A}(x_a)$ is the helicity 
density matrix of parton $a$ inside the polarized hadron $A$; similarly
for $\rho_{\la^{\,}_b, \la^{\prime}_b}^{b/B,S_B}(x_b)$.
The $\hat M_{\la^{\,}_c, \la^{\,}_d; \la^{\,}_a, \la^{\,}_b}$'s
are the helicity amplitudes for the elementary process $ab \to cd$;
if one wishes to consider higher order (in $\alpha_s$) contributions also
elementary processes involving more partons should be included. 
$D_{\la^{\,}_C,\la^{\prime}_C}^{\la^{\,}_c,\la^{\prime}_c}(z)$ is the 
product of {\it fragmentation amplitudes} for the $c \to C + X$ process
\beq
D_{\la^{\,}_C,\la^{\prime}_C}^{\la^{\,}_c,\la^{\prime}_c} 
= \> \sumint_{X, \la_{X}} {\cal D}_{\la^{\,}_{X},\la^{\,}_C;
\la^{\,}_c} \, {\cal D}^*_{\la^{\,}_{X},\la^{\prime}_C; \la^{\prime}_c}
\, ,
\label{framp}
\eeq
where the $\sumint_{X, \la_{X}}$ stands for a spin sum and phase space
integration of the undetected particles, considered as a system $X$.
The usual unpolarized fragmentation function $D_{C/c}(z)$, {\it i.e.} 
the density number of hadrons $C$ resulting from the fragmentation of 
an unpolarized parton $c$ and carrying a fraction $z$ of its momentum,
is given by
\beq
D_{C/c}(z) = {1\over 2} \sum_{\la^{\,}_c,\la^{\,}_C}  
D_{\la^{\,}_C,\la^{\,}_C}^{\la^{\,}_c,\la^{\,}_c}(z)
\,. \label{fr}
\eeq
For simplicity of notations we have not shown in Eq. (\ref{dsabpol}) the $Q^2$ 
scale dependences in the unpolarized distribution and fragmentation functions,
$f$ and $D$.

In the case in which only one hadron, say $A$, is polarized (orthogonally to 
the scattering plane) Eq. (\ref{dsabpol}) reads
\barr
\frac{E_C \, d^3\sigma^{A^\uparrow B \to C X}} {d^{3} \bfp_C} &=&
\sum_{a,b,c,d;\{\lambda\}}
\rho_{\la^{\,}_a, \la^{\prime}_a}^{a/A^\uparrow} \, f_{a/A}(x_a) \,
\otimes \frac12 \, f_{b/B}(x_b) 
\label{dsapol} \\ 
&\otimes& \hat M_{\la^{\,}_c, \la^{\,}_d; \la^{\,}_a, \la^{\,}_b} \,
\hat M^*_{\la^{\prime}_c, \la^{\,}_d; \la^{\prime}_a, \la^{\,}_b} \,
\otimes D_{\la^{\,}_C,\la^{\,}_C}^{\la^{\,}_c,\la^{\prime}_c}(z) \>. 
\nonumber 
\earr

Eqs. (\ref{dsabpol}) and (\ref{dsapol}) hold at leading twist and large $p_T$
values of the produced hadron; the intrinsic $\bfk_\perp$ of the partons
have been integrated over and collinear configurations dominate both the 
distribution functions and the fragmentation processes; one can then see that,
in this case, there cannot be any single spin asymmetry. In fact, total 
angular momentum conservation in the (forward) fragmentation process [see 
Eq. (\ref{framp})] implies $\la^{\,}_c = \la^{\prime}_c$; this, in turns,
together with helicity conservation in the elementary process, implies
$\la^{\,}_a = \la^{\prime}_a$. If we further notice that, by parity
invariance, 
$\sum_{\la^{\,}_C} D_{\la^{\,}_C,\la^{\,}_C}^{\la^{\,}_c,\la^{\,}_c}$ does
not depend on $\la^{\,}_c$ and that 
$\sum_{\la^{\,}_b,\la^{\,}_c,\la^{\,}_d} 
|\hat M_{\la^{\,}_c, \la^{\,}_d; \la^{\,}_a, \la^{\,}_b}|^2$ 
is independent of $\la^{\,}_a$, we remain with $\sum_{\la^{\,}_a}  
\rho^{a/A^\uparrow}_{\la^{\,}_a, \la^{\,}_a} = 1$. Moreover, in the 
absence of intrinsic $\bfk_\perp$ and initial state interactions,
the parton density numbers $f_{a/A}(x_a)$ cannot depend on the spin of $A$
and any spin dependence disappears from Eq. (\ref{dsapol}), so that:
\beq
d\sigma^{A^\uparrow B \to C X} - d\sigma^{A^\downarrow B \to C X} = 0 \>.
\label{an0}
\eeq
Therefore one does not expect any sizeable single spin (or left-right) 
asymmetry:
\beq
A_N \equiv \frac{d\sigma^\uparrow(\bfp_T) - d\sigma^\downarrow(\bfp_T)}
{d\sigma^\uparrow(\bfp_T) + d\sigma^\downarrow(\bfp_T)} =
\frac{d\sigma^\uparrow(\bfp_T) - d\sigma^\uparrow(-\bfp_T)}
{2\,d\sigma^{unp}} \> \cdot \label{an}
\eeq 
This is contradicted by data \cite{e704} showing large values of $A_N$
in $\pup p \to \pi X$ and $\bar p^\uparrow p \to \pi X$ processes. 

\vskip 12pt
\noindent
{\bf Intrinsic $\bfk_\perp$ effects in fragmentation and/or distribution
functions: {\mbox{\boldmath $A_N \not= 0$}}}
\vskip 6pt

The above conclusion may be avoided by considering the transverse
motion of the quarks relatively to the parent hadron or of the
observed hadron relatively to the fragmenting quark. 
The original suggestion that the intrinsic $\bfk_\perp$ of the quarks in
the distribution functions might give origin to single spin asymmetries
was first made by Sivers \cite{siv}; such an effect is not forbidden
by QCD time reversal invariance \cite{col1} provided one takes into account 
soft initial state interactions among the colliding hadrons \cite{noi2}. 
A similar suggestion for a possible origin of single spin asymmetries
was later made by Collins \cite{col1}, concerning 
transverse momentum effects in the fragmentation of a polarized quark. 
A consistent phenomenological application of these ideas can be found
in a series of papers \cite{noi1,noi2,noi3}. More recently, a further 
possible source of single spin effects, related to  
distribution functions, was discussed by Boer \cite{dan}. 

When allowing for parton intrinsic motion spin effects can remain in
new -- spin and $\bfk_\perp$ dependent -- distribution or fragmentation
functions. We list these functions in the sequel.

$\Delta^N f_{q/\pup}(x, \bfk_{\perp})$ \cite{siv} is the difference 
between the density numbers $\hat f_{q/\pup}(x, \bfk_{\perp})$ and 
$\hat f_{q/\pdown}(x, \bfk_{\perp})$ of quarks $q$, with all possible 
polarization, longitudinal momentum fraction $x$ and intrinsic transverse 
momentum $\bfk_{\perp}$, inside a transversely polarized proton with spin 
$\uparrow$ or $\downarrow$: 
\barr
\Delta^Nf_{q/\pup}(x, \bfk_{\perp})  &\equiv& 
\hat f_{q/\pup}(x, \bfk_{\perp})-\hat f_{q/\pdown}(x, \bfk_{\perp}) 
\label{delf1}\\
& = & \hat f_{q/\pup}(x, \bfk_{\perp})-\hat f_{q/\pup}(x, - \bfk_{\perp}) 
\nonumber
\earr
where the second line follows from the first one by rotational invariance.
Notice that $\Delta^Nf_{q/\pup}(x,\bfk_{\perp})$ 
vanishes when $\bfk_{\perp} \to 0$; parity invariance also requires 
$\Delta^Nf$ to vanish when the proton transverse spin has no component
perpendicular to $\bfk_{\perp}$, so that
\beq
\Delta^Nf_{q/\pup}(x,\bfk_{\perp}) \sim k_{\perp} \, \sin\alpha
\label{sinf}
\eeq
where $\alpha$ is the angle between $\bfk_{\perp}$ and the $\uparrow$ 
direction.

$\Delta^N f$ by itself is a leading twist distribution function, but
its $\bfk_\perp$ dependence, when convoluted with the 
elementary partonic cross-section, results in twist-3 contributions
to single spin asymmetries. This same function (up to some factors)
has also been introduced in Ref. \cite{mul1} -- where it is 
denoted by $f_{1T}^\perp$ -- as a leading twist $T$-odd distribution 
function. The exact relation between 
$\Delta^N f$ and $f_{1T}^\perp$ is discussed in Ref. \cite{bm}.

A function analogous to $\Delta^Nf_{q/\pup}(x, \bfk_{\perp})$
can be defined for the fragmentation process of a transversely 
polarized parton \cite{col1}, giving the difference between the 
density numbers $\hat D_{h/\qup}(z, \bfk_{\perp})$ and 
$\hat D_{h/\qdown}(z, \bfk_{\perp})$
of hadrons $h$, with longitudinal momentum fraction $z$ and transverse 
momentum $\bfk_{\perp}$ inside a jet originated by the fragmentation of a 
transversely polarized quark with spin $\uparrow$ or $\downarrow$:
\barr
\Delta^N D_{h/\qup}(z, \bfk_{\perp}) &\equiv&
\hat D_{h/\qup}(z, \bfk_{\perp}) - \hat D_{h/\qdown}(z, \bfk_{\perp}) 
\label{deld1}\\
&=& \hat D_{h/\qup}(z, \bfk_{\perp})-\hat D_{h/\qup}(z, - \bfk_{\perp}) \>.
\nonumber
\earr
A closely related function is denoted by $H_1^\perp$ in 
Refs. \cite{mul1, mul2} and its correspondence with $\Delta^ND$ 
is discussed in Ref. \cite{bm}. Again we expect
\beq
\Delta^ND_{h/\qup}(z,\bfk_{\perp}) 
\sim k_{\perp} \, \sin\beta
\label{sind}
\eeq
where $\beta$ is the angle between $\bfk_{\perp}$ and the $\uparrow$ 
direction.

Similarly one can introduce the difference
$\Delta^N f_{\qup/p}(x, \bfk_{\perp})$  
between the density numbers $\hat f_{\qup/p}(x, \bfk_{\perp})$ and 
$\hat f_{\qdown/p}(x, \bfk_{\perp})$ of quarks $q$, with spin 
$\uparrow$ or $\downarrow$, longitudinal momentum fraction $x$ and 
intrinsic transverse momentum $\bfk_{\perp}$, inside an unpolarized proton: 
\barr
\Delta^Nf_{\qup/p}(x, \bfk_{\perp})  &\equiv& 
\hat f_{\qup/p}(x, \bfk_{\perp})-\hat f_{\qdown/p}(x, \bfk_{\perp}) 
\label{delf2}\\
& = & \hat f_{\qup/p}(x, \bfk_{\perp})-\hat f_{\qup/p}(x, - \bfk_{\perp}) \>. 
\nonumber
\earr
A closely related function, denoted by $h_1^\perp$, was discussed
in Ref. \cite{dan}.

Finally, although not relevant for the single spin asymmetries we 
consider here, one could introduce the difference between the 
density numbers $\hat D_{\hup/q}(z, \bfk_{\perp})$ and 
$\hat D_{\hdown/q}(z, \bfk_{\perp})$
of hadrons $h$, with longitudinal momentum fraction $z$ and transverse 
momentum $\bfk_{\perp}$ inside a jet originated by the fragmentation of 
an unpolarized quark $q$:
\barr
\Delta^N D_{\hup/q}(z, \bfk_{\perp}) &\equiv&
\hat D_{\hup/q}(z, \bfk_{\perp}) - \hat D_{\hdown/q}(z, \bfk_{\perp}) 
\label{deld2}\\
&=& \hat D_{\hup/q}(z, \bfk_{\perp})-\hat D_{\hdown/q}(z, - \bfk_{\perp}) 
\>. \nonumber
\earr
Such a function might prove useful in tackling the longstanding 
problem of hyperon polarization in inclusive $p-N$ processes \cite{prep}. 
A closely related function is denoted by $D_{1T}^\perp$ in 
Refs. \cite{mul1, mul2}. 

Assuming that the factorization theorem, Eq. (\ref{dsapol}), holds 
when parton intrinsic motion is taken into account, and using the new 
functions (\ref{delf1}), (\ref{deld1}) and (\ref{delf2}), one has, at 
leading order in $k_\perp$, for the $\pup p \to \pi X$ process:
\barr
\frac{E_\pi \, d^3\sigma^\uparrow} {d^{3} \bfp_\pi} 
- \frac{E_\pi \, d^3\sigma^\downarrow} {d^{3} \bfp_\pi} &=&
\sum_{a,b,c,d} \int \frac {dx_a \, dx_b} {\pi z} \> \times 
\Bigg\{ \label{gen} \\
&\,& \!\!\! \int d^2 \bfk_{\perp} \,
\Delta^Nf_{a/\pup} (x_a, \bfk_{\perp}) \> f_{b/p}(x_b) \,
\frac{d \hat \sigma} {d\hat t} (x_a, x_b, \bfk_{\perp}) \>
D_{\pi/c}(z) \nonumber \\
&+& \!\!\! \int d^2 \bfk'_{\perp}\, h_1^{a/p}(x_a) \> f_{b/p}(x_b) \>
\Delta_{NN} \hat\sigma(x_a, x_b, \bfk'_\perp) \>
\Delta^N D_{\pi/c}(z, \bfk'_\perp) \nonumber \\  
&+& \!\!\! \int d^2 \bfk''_{\perp}\, h_1^{a/p}(x_a) \> 
\Delta^Nf_{\bup/p} (x_b, \bfk''_{\perp}) \>
\Delta'_{NN} \hat\sigma(x_a, x_b, \bfk''_\perp) \>
D_{\pi/c}(z) \Bigg\}, \nonumber
\earr
where the second line corresponds to the so-called Sivers effect \cite{siv}, 
the third to Collins effect \cite{col1} and the fourth one to the mechanism 
recently proposed by Boer \cite{dan}. Apart from the new functions 
$\Delta^Nf$ and $\Delta^ND$, the other quantities appearing in Eq. (\ref{gen})
are the unpolarized quark distribution and fragmentation functions, $f$ and 
$D$; the unpolarized cross-section for the elementary process $ab \to cd$,
$d\hat{\sigma}/d\hat{t}$; the transverse spin content of the proton:
\beq
 h_1^{q/p} = f_{\qup/\pup}(x) - f_{\qdown/\pup}(x)
\eeq
and the elementary double spin asymmetries, computable in pQCD:
\beq 
\Delta_{NN} \hat\sigma = {d\hat \sigma^{\aup b \to \cp d} \over d\hat t} 
- {d\hat \sigma^{\aup b \to \cdown d} \over d\hat t} \,, \label{dnn1}
\eeq
\beq 
\Delta'_{NN} \hat\sigma = {d\hat \sigma^{\aup \bup \to c d} \over d\hat t} 
- {d\hat \sigma^{\aup \bdown \to c d} \over d\hat t} \,\cdot \label{dnn2}
\eeq

An equation with the same structure and a similar physical meaning as
Eq. (\ref{gen}) can be found in Ref. \cite{qs}, where the new functions 
$\Delta^Nf$ and $\Delta^ND$ are replaced by higher twist parton correlation 
functions. 

Eq. (\ref{gen}), or some of its terms, can be used for a phenomenological
description of single spin asymmetries. We next summarize what has been done.

\vskip 12pt
\noindent
{\bf Fits to existing data, predictions}
\vskip 6pt
\noindent
{\it Sivers effect only}
\vskip 4pt 
In Refs. \cite{noi2} and \cite{noi3} the scheme of Eq. (\ref{gen}) was
adopted, taking into account only Sivers effect; to simplify things 
it was assumed -- in this very first application of the idea of intrinsic
quark motion and spin dependence -- that the integral over $\bfk_\perp$
is dominated by configurations in which $\bfk_\perp$ lies in the
scattering plane [that is, $\sin\alpha = 1$ in Eq. (\ref{sinf})] and
its magnitude equals some average value \cite{noi1}:
\beq
\frac{1}{M} \, k_{\perp}^0(x_a) =
 0.47 \> x_a^{0.68}(1-x_a)^{0.48} \,,
\label{kx}
\eeq
where  $M$ = 1 GeV/$c^2$. 

The residual $x_a$ dependence in $\Delta^Nf_{a/\pup}$ not coming from
$k_{\perp}^0$ was taken to be of the simple form 

\beq
N_a \, x_a^{\alpha_a}(1-x_a)^{\beta_a} \,,
\label{ixf}
\eeq
where $N_a$, $\alpha_a$ and $\beta_a$ are free parameters. Only $u$ and
$d$ quark contributions to $\Delta^Nf_{a/\pup}$ were considered. 

One then ends up with the simple expression:
\barr
&&\int d^2\bfk_{\perp} \Delta^Nf_{a/\pup}(x_a,\bfk_{\perp}) 
 \left[ \frac{d\hat\sigma}{d\hat t}(\bfk_{\perp}) -
 \frac{d\hat\sigma}{d\hat t}(-\bfk_{\perp}) \right] \nonumber \\
&\simeq& \frac{k_{\perp}^0(x_a)}{M} \> N_a \, x_a^{\alpha_a}(1-x_a)^{\beta_a}
 \left[ \frac{d\hat\sigma}{d\hat t}(\bfk_{\perp}^0) -
 \frac{d\hat\sigma}{d\hat t}(-\bfk_{\perp}^0) \right] \,,
\label{app}
\earr
where $k_{\perp}^0(x_a)$ is given by Eq. (\ref{kx}) and, choosing 
$xz$ as the scattering plane and $z$ as the direction of the incoming
polarized proton, $\bfk_{\perp}^0 = (k_{\perp}^0,0,0)$.

Inserting this result into the first two lines of Eq. (\ref{gen}),
one can compute the single spin asymmetry (\ref{an}) in terms of the
parameters $\alpha_a, \> \beta_a, \> N_a$; only the leading valence quark 
contributions to $\Delta^Nf$ were considered in the numerator of $A_N$, 
while all leading 
order pQCD elementary processes were included in the denominator. The 
parameters were fixed by performing a best fit to the experimental data, 
with the results shown in Fig. 1, corresponding to the $\Delta^Nf$ functions:
\beq
\Delta^Nf_{u/\pup}(x, k_\perp^0) = 6.92 \, x^{2.02} (1-x)^{4.06}  \quad\quad
\Delta^Nf_{d/\pup}(x, k_\perp^0) = -2.33 \, x^{1.44} (1-x)^{4.62}
\label{dffit}
\eeq

It is clear from Fig. 1 how Sivers effect alone allows a good fit of
the experimental data. Let us add a few more comments:
\begin{itemize}

\item
The resulting expressions of $\Delta^Nf_{q/\pup}$, Eq. (\ref{dffit}), are 
reasonable and quite acceptable; in particular they satisfy the
positivity condition $|\Delta^Nf_{q/\pup}| \leq 2 \, f_{q/p}\,$.

\item
The opposite sign of $\Delta^Nf_{u/\pup}$ and $\Delta^Nf_{d/\pup}$
is expected from transverse momentum conservation.

\item
It is easy to explain, within this scheme, why almost opposite values
of $A_N^{\pi^+}$ and $A_N^{\pi^-}$ do not imply, as one might naively
expect,  $A_N^{\pi^0} \simeq 0$ \cite{noi2}.

\item
This same scheme, with the same $\Delta^Nf$ functions (\ref{dffit}), was 
used -- without any free parameter -- to compute $A_N$ in $\bar p^\uparrow p 
\to \pi X$; the agreement with data \cite{e704} is good \cite{noi3}.
\end{itemize}
 
\vskip 6pt
\noindent
{\it Collins effect only}
\vskip 4pt
A similar analysis was performed in Ref. \cite{noi1} taking into account 
only Collins effect. Again, it was assumed that the main contribution comes
from an average $k_\perp^0$ value, with a simple parametrization of 
$\Delta^ND_{\pi/c}(z, k_\perp^0)$ for all leading valence quarks:
\beq
\Delta^ND_{\pi/c}(z, k_\perp^0) = \frac{k_{\perp}^0(z)}{M} \,
N_c \, z^{\alpha_c}\,(1-z)^{\beta_c} \,,
\label{ixd}
\eeq
where $N_c$, $\alpha_c$ and $\beta_c$ are free parameters and
\beq
\frac{k_{\perp}^0(z)}{M} = 0.61 \; z^{0.27} \; (1-z)^{0.20}\;,
\label{kfit}
\eeq
with $M$ = 1 GeV/$c^2$.

Another free parameter is contained in the expression of
$h_1^{q/p}(x) = P^{q/\pup}\,f_{q/p}(x)$ where $P^{q/\pup}$ is
the transverse polarization of quark $q$ inside the transversely
polarized proton. 

The experimental data \cite{e704} were best fitted as in Fig. 2, with the 
resulting $\Delta^ND$ function and $P^{q/\pup}$ values \cite{noi1}:
\barr
\Delta^ND_{\pi/q}(z, k_\perp^0) &=& -0.13 \, z^{2.60}\;(1-z)^{0.44} 
\quad\quad z \le 0.977 \nonumber \\
\Delta^ND_{\pi/q}(z, k_\perp^0) = - 2 D_{\pi/q}(z) &=& - 2.20 \; z^{-1} \; 
(1-z)^{1.2} \quad\quad\>\>\> z > 0.977  \label{bfit2} \\
P^{u/\pup} = \frac{2}{3} && P^{d/\pup} = -0.88 \>. \nonumber
\earr

The quality of the fit is comparable to that obtained using only Sivers
effect, shown in Fig. 1. However, some comments are now necessary:

\begin{itemize}
\item
In order to fit the data, $\Delta^ND_{\pi/q}$ has to saturate at large $z$
the positivity constraint $|\Delta^ND_{\pi/q}| \le 2\,D_{\pi/q}$.
Otherwise, the values of $A_N$ at large $x_F$ would be much too small.

\item
The resulting value of $h_1^{d/q}$ is
\beq
h_1^{d/p}(x) = -0.88 \, f_{d/p}(x) \>, \label{h1d}
\eeq
which violates the Soffer's bound \cite{sof}
\beq
|h_1^{d/p}| \le \frac12 (f_{d/p} + \Delta d) \>. \label{sofb}
\eeq
$\Delta d$ is the $d$ quark helicity distribution, which, being
in most parametrization of data negative, makes bound (\ref{sofb})
very strict. 

\item
A parametrization of $\Delta^ND$ which satisfies Soffer's bound was
used in Ref. \cite{bl}, with a good resulting fit, provided one allows
$\Delta d$ to become positive at large $x$ values; however, also in 
this case the positivity constraint has to be saturated at large $z$. 

\end{itemize}

We can at this point conclude that the phenomenological approach to 
single spin asymmetries based on the generalization of the factorization
theorem with the inclusion of quark intrinsic motion and on the 
introduction of the new spin and $\bfk_\perp$ dependent functions
(\ref{delf1}), (\ref{deld1}) and (\ref{delf2}) is indeed promising
and worth being pursued. The relative contributions of the several
terms in Eq. (\ref{gen}) is still unknown; a first guess would indicate
that Sivers effect is indeed necessary, while Collins effect alone
leads to unreasonably large values of $|\Delta^ND|$, the Collins function.
The third effect, originating from the quarks in the unpolarized hadron
has not been studied yet, although it should not be relevant
for large and positive $x_F$ values. We turn now to a study of
Collins effect in DIS.   

\goodbreak

\vskip 12pt
\noindent
{\bf Fragmentation of a polarized quark in semi-inclusive DIS}
\vskip 6pt
\nobreak
The inclusive production of hadrons in DIS with transversely polarized
nucleons, $\ell N^\uparrow \to \ell h X$,
is the ideal process to study Collins effect; in such a case, in fact,
Sivers effect, which requires initial state interactions \cite{noi1, noi4},
is negligible and any single spin asymmetry must originate from 
spin dependences in the fragmentation of a polarized quark.

If one looks at the $\gamma^* N^\uparrow \to h X$ process in the $\gamma^*-N$
c.m. system, the elementary interaction is simply a $\gamma^*$ hitting a 
transversely polarized quark, which bounces back and fragments into
a jet containing the detected hadron. The hadron $p_T$ in this case 
coincides with its $k_\perp$ inside the jet; the fragmenting quark 
polarization can be computed from the initial quark one. 

The spin and $\bfk_\perp$ dependent fragmentation function for a quark with 
momentum $\bfp_q$ and a {\it transverse} polarization vector $\bfP_q$ 
($\bfp_q \cdot \bfP_q = 0$) which fragments into a hadron with momentum 
$\bfp_h = z\bfp_q + \bfp_T$ ($\bfp_q \cdot \bfp_T = 0$) can be written as:
\beq 
D_{h/q}(\bfp_q, \bfP_q; z, \bfp_T) = \hat D_{h/q}(z, p_T) + \frac 12 \> 
\Delta^ND_{h/q}(z, p_T) \> \frac{\bfP_q \cdot (\bfp_q \times \bfp_T)}
{|\bfp_q \times \bfp_T|} \label{colfn}
\eeq
where $\hat D_{h/q}(z, p_T)$ is the unpolarized fragmentation function and
$\Delta^ND$ is the same function as introduced in Eq. (\ref{deld1}).
Notice that -- as required by parity invariance -- the only component 
of the polarization vector which contributes to the spin dependent part
of $D$ is that perpendicular to the $q-h$ plane; in general one has:
\beq
\bfP_q \cdot 
\frac{\bfp_q \times \bfp_T} {|\bfp_q \times \bfp_T|}
= P_q \sin\Phi_C \>, \label{colan}
\eeq
where $P_q = |\bfP_q|$ and we have defined the {\it Collins angle} $\Phi_C$. 

Eq. (\ref{colfn}) leads to a possible single spin asymmetry in 
semi-inclusive DIS off nucleons with polarization $\pm\bfP$,
\beq  
A^h_N \equiv
 \frac{d\sigma^{\ell + p,\bfP \to \ell' + h + X}
      -d\sigma^{\ell + p,-\bfP \to \ell' + h + X}}
      {d\sigma^{\ell + p,\bfP \to \ell' + h + X}
      +d\sigma^{\ell + p,-\bfP \to \ell' + h + X}} \>, \label{asym1}
\eeq
which is given by \cite{noi5}
\beq
A^h_N(x,y,z,\Phi_C, p_T) =
\frac{\sum_q e_q^2 \, h_1^{q/p}(x) \> \Delta^ND_{h/q}(z, p_T)}
{2\sum_q e_q^2 \, f_{q/p}(x) \> D_{h/q}(z, p_T)} \>
\frac{2(1-y)} {1 + (1-y)^2} \> P_T \> \sin\Phi_C \>, \label{asym2} 
\eeq
where $P_T$ is the nucleon polarization vector component {\it transverse}
with respect to the $\gamma^*$ direction; $x$ and $y$ are the usual DIS
variables. If one collects data at different kinematical values one should
integrate over the relevant $x$ and $y$ regions so that Eq. (\ref{asym2}) 
reads:
\beq
A^h_N(z, \Phi_C, p_T) =
\frac{\sum_q \int dx \, dy \, e_q^2 \, h_{1q}(x) \> 2(1-y)/(xy^2) \> 
\Delta^ND_{h/q}(z, p_T) \> P_T \> \sin\Phi_C}
{2\sum_q \int dx \, dy \, e_q^2 \, f_{q/p}(x) \> (1 + (1-y)^2)/(xy^2) \>
D_{h/q}(z, p_T)} \> \cdot \label{asint}  
\eeq

Some preliminary data on $A_N^{\pi}$ have recently appeared \cite{her, smc} 
and have been analysed in Ref. \cite{noi5} using Eq. (\ref{asym2}).
By saturating the unknown values of $h_1^{u,d/p}$ with the Soffer's bound 
[see Eq. (\ref{sofb})] lower bounds for the Collins function have been 
obtained for the fragmentation of a $u$ quark into a $\pi^+$. From
SMC data \cite{smc} one has 
\beq
\frac {|\Delta^ND_{\pi/q}(\langle z \rangle, \langle p_T \rangle)|} 
{2\,D_{\pi/q}(\langle z \rangle, \langle p_T \rangle)} \> 
\gsim \> (0.24 \pm 0.15)
\quad\quad \langle z \rangle \simeq 0.45 \>, \quad 
\langle p_T \rangle \simeq 0.65 \> \mbox{GeV}/c \>,
\label{res2}
\eeq  
and from HERMES data \cite{her}
\beq
\frac {|\Delta^ND_{\pi/q}(z, p_T)|} {2\,D_{\pi/q}(z, p_T)} \>
\gsim  \> 0.20 \pm 0.04(stat.) \pm 0.04(syst.) 
\quad\quad\quad  z \geq 0.2 \>. \label{res3}
\eeq

If confirmed, HERMES and SMC data indicate a large value of the Collins
function, which might then play a significant role in other processes.
In particular, it would be of great interest to compare the single spin
asymmetry measured in the inclusive process $\ell \pup \to \pi X$
with that measured by E704 Collaboration \cite{e704} in $\pup p \to \pi X$
processes: if the origin of the asymmetry is mainly in Collins mechanism
similar results should be found in both cases. More data on single transverse
spin asymmetries will be available in the future from operating or 
progressing facilities like JLAB, RHIC and COMPASS.   

\vskip 6pt
{\bf Acknowledgements}
\vskip 4pt
\noindent
One of us (M.A.) would like to thank the organizers of the Workshop for the 
kind invitation.

\newpage

%\vskip -4pt

%\end{document}
%%%%%%%%%%%%%%%%%%%%%%%%%%%%%%%%%%%%%%%%%%%%%%%%%%%%%%%%%%%%%%%%%%%%%%%%%%%%%%%
\newpage
%%%%%%%%%%%%%%%%%%%%%%%%%%%%%%%%%%%%%%%%%%%%%%%%%%%%%%%%%%%%%%%%%%%%%%%%%%
\begin{figure}[t]
\begin{center}
\mbox{~\epsfig{file=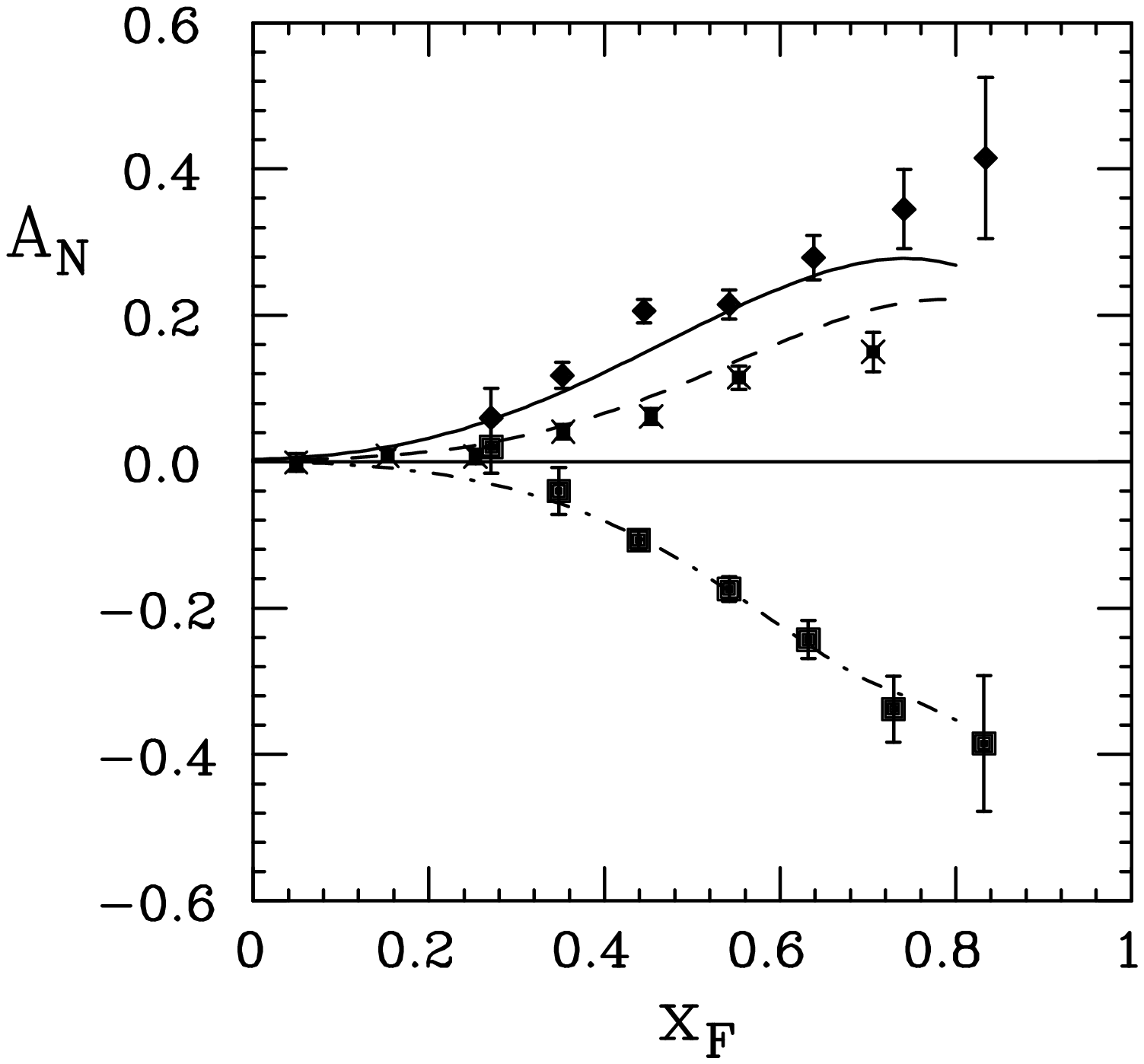,angle=0,width=8.5cm}}
\vspace{-2.5cm}
\caption{\small
Fit of the data on $A_N$ for the process $p^\uparrow p\to\pi\,X$
\protect\cite{e704}, as obtained in Ref. [7], assuming that only Sivers
effect is active; the upper, middle, and lower sets of data and curves
refer respectively to $\pi^+$, $\pi^0$, and $\pi^-$.}
\end{center}
\end{figure}
%%%%%%%%%%%%%%%%%%%%%%%%%%%%%%%%%%%%%%%%%%%%%%%%%%%%%%%%%%%%%%%%%%%%%%%%%%%%%%%
%\newpage
%%%%%%%%%%%%%%%%%%%%%%%%%%%%%%%%%%%%%%%%%%%%%%%%%%%%%%%%%%%%%%%%%%%%%%%%%%
\begin{figure}[c]
\begin{center}
\hspace{-20pt}
\mbox{~\epsfig{file=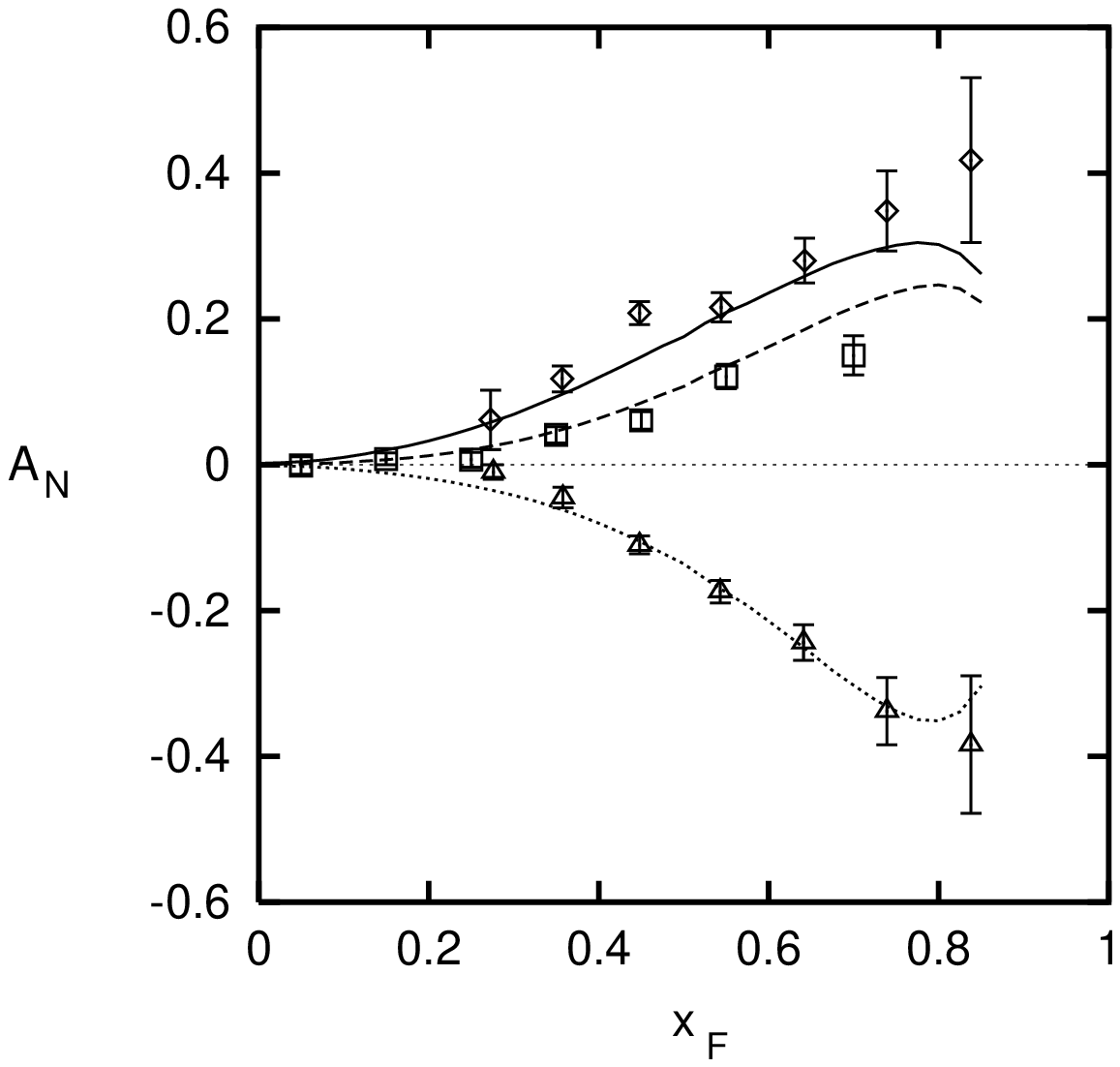,angle=-90,width=7.2cm}}
\vspace{0.5cm}
\caption{ \small
Fit of the data on $A_N$ for the process $p^\uparrow p\to\pi\,X$
\protect\cite{e704}, as obtained in Ref. [3], assuming that only Collins
effect is active; the upper, middle, and lower sets of data and curves
refer respectively to $\pi^+$, $\pi^0$, and $\pi^-$.}
\end{center}
\end{figure}
%%%%%%%%%%%%%%%%%%%%%%%%%%%%%%%%%%%%%%%%%%%%%%%%%%%%%%%%%%%%%%%%%%%%%%%%%%%
\end{document}